\documentclass[conference]{IEEEtran}
\IEEEoverridecommandlockouts

\usepackage{cite}
\usepackage{amsmath,amssymb,amsfonts}
\usepackage{graphicx}
\usepackage{booktabs}
\usepackage{multirow}
\usepackage{xcolor}
\usepackage{url}
\usepackage{tikz}
\usetikzlibrary{positioning,arrows.meta,fit,backgrounds,calc}

\graphicspath{{./figures/}{./}}

\newcommand{\ariarepo}{\url{https://github.com/matteospanio/aria}}

\begin{document}

\title{A Quantized Native Runtime for On-Device Semantic Audio Generation}

\author{\IEEEauthorblockN{Matteo Spanio}
\IEEEauthorblockA{\textit{Centro di Sonologia Computazionale (CSC)} \\
\textit{Dept.\ of Information Engineering, University of Padova}\\
Padova, Italy \\
spanio@dei.unipd.it}
\and
\IEEEauthorblockN{Antonio Rod\`a}
\IEEEauthorblockA{\textit{Centro di Sonologia Computazionale (CSC)} \\
\textit{Dept.\ of Information Engineering, University of Padova}\\
Padova, Italy \\
roda@dei.unipd.it}
}

\maketitle

\begin{abstract}
Semantic audio applications increasingly require controllable generation on commodity and embedded hardware rather than through framework-heavy datacenter stacks. We present \textit{aria}, a dependency-free native runtime that runs the complete text-to-music pipeline of Stable Audio~3 (SA3) on ordinary GPUs, CPU-only machines, and a Raspberry~Pi~5, with no Python or deep-learning framework underneath. Our main contribution is a study of \emph{quantization}: running the model at lower numerical precision to fit tight memory budgets, saving memory in place rather than adding to it. Because the runtime owns every internal tensor, it also exposes activation steering, a low-cost way to steer what the model generates. We judge the quality cost with three independent measures of the output (prompt adherence, overall audio quality, taste preservation), each compared against the ordinary variation between random seeds. Eight-bit precision shows no measurable quality loss on any measure while sharply cutting memory, and it is the fastest mode on the GPU; four-bit adds a small, bounded cost but shrinks the footprint enough to run the $1.2$-billion-parameter model on an $8$\,GB Pi. Against the official implementation, aria matches or exceeds generation speed and starts about seven times faster. A case study of the steering interface generates music carrying taste associations (\emph{sonic seasoning}), with genuine but bounded control for a subset of attributes. These results make a compact, quantized runtime with built-in control a practical basis for on-device semantic audio in Internet-of-Sounds settings. The \textit{aria} runtime is released at \ariarepo.
\end{abstract}


\section{Introduction}
\label{sec:intro}
Semantic audio generation is moving from cloud demos toward interactive tools,
local creative services, and embedded audio devices~\cite{evans2025icassp}. These settings care not only
about generation quality but also about deployment properties: cold-start
latency, predictable memory use, portability across commodity CPUs and GPUs, and
the ability to keep a model resident near the user. Open-weight music models are
now strong enough to be useful in such systems, yet their reference
implementations still assume a Python/PyTorch serving stack and a discrete GPU.
That mismatch matters for Internet-of-Sounds applications, where semantic audio
often needs to run as a local or edge service rather than as a heavyweight
datacenter job.

\begin{figure}[!t]
  \definecolor{cbad}{RGB}{213,94,0}
  \definecolor{cgood}{RGB}{0,158,115}
  \definecolor{cacc}{RGB}{0,114,178}
  \centering
  \begin{tikzpicture}[
    font=\scriptsize,
    panel/.style={draw=#1!55, rounded corners=3pt, fill=#1!4, line width=0.6pt,
                  inner sep=0pt},
    mbox/.style={draw=#1!60, rounded corners=2pt, fill=#1!10, align=center,
                 inner sep=3pt, line width=0.6pt},
    badge/.style={draw=#1!60, rounded corners=2.5pt, fill=#1!12, align=center,
                  inner sep=2.4pt, line width=0.6pt, minimum width=26mm},
    chip/.style={draw=cacc!60, rounded corners=5pt, fill=cacc!10, inner sep=2.8pt,
                 line width=0.5pt},
    blk/.style={draw=black!50, fill=white, minimum width=3.1mm, minimum height=5mm,
                line width=0.45pt},
    site/.style={circle, inner sep=1.35pt, font=\bfseries\tiny, text=white},
    >={Stealth[length=1.8mm]},
  ]
    \node[mbox=cgood, inner sep=3pt] (pr) at (-3.55, -3.95) {prompt};
    \node[mbox=cgood, inner sep=3pt, right=5mm of pr] (t5) {T5};
    \node[blk, minimum width=4.2mm, right=6mm of t5] (k1) {};
    \node[blk, minimum width=4.2mm, right=0.9mm of k1] (k2) {};
    \node[blk, minimum width=4.2mm, right=0.9mm of k2, fill=cbad!25] (k3) {};
    \node[blk, minimum width=4.2mm, right=0.9mm of k3] (k4) {};
    \node[mbox=cgood, inner sep=3pt, right=6mm of k4] (de) {decoder};
    \node[mbox=cgood, inner sep=3pt, right=5mm of de] (au) {audio};
    \draw[->] (pr) -- (t5); \draw[->] (t5) -- (k1);
    \draw[->] (k4) -- (de); \draw[->] (de) -- (au);
    \node[site, fill=cbad]  (s1) at ($(k3.north)+(0,2.6mm)$) {1};
    \node[site, fill=cgood] (s2) at ($(k4.east)!0.5!(de.west)+(0,-3.4mm)$) {2};
    \node[site, fill=cacc]  (s3) at ($(t5.east)!0.5!(k1.west)+(0,-3.4mm)$) {3};
    \draw[->, cbad,  line width=0.7pt] (s1) -- (k3.north);
    \draw[->, cgood, line width=0.7pt] (s2) -- ($(k4.east)!0.5!(de.west)$);
    \draw[->, cacc,  line width=0.7pt] (s3) -- ($(t5.east)!0.5!(k1.west)$);
    \node[font=\tiny, text=black!60] at ($(k1.west)!0.5!(k4.east)+(0,-4.1mm)$) {DiT blocks};
    \node[badge=cgood, minimum width=23mm, anchor=west] (gb1) at (pr.west |- 0,-5.35)
        {\textbf{1.6--2.9\,s} cold, $7\times$};
    \node[badge=cgood, minimum width=23mm, anchor=east] (gb3) at (au.east |- 0,-5.35)
        {\textbf{Pi 5}: both models\;\raisebox{-0.7mm}{\includegraphics[height=3.2mm]{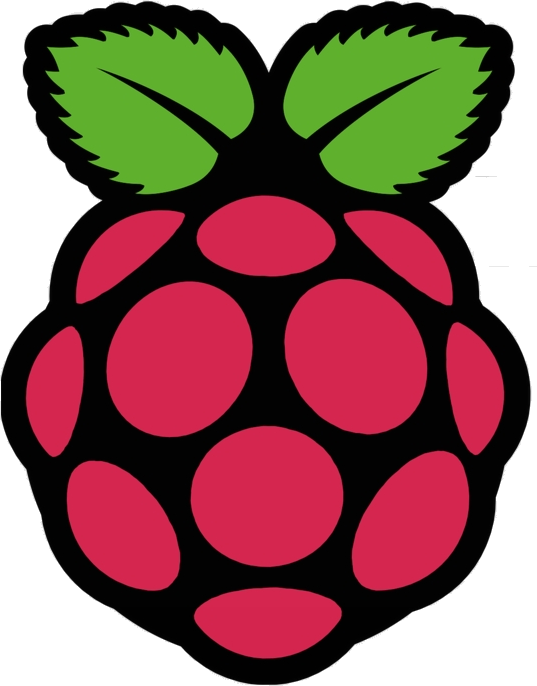}}};
    \node[badge=cgood, minimum width=23mm] (gb2) at ($(gb1.east)!0.5!(gb3.west)$)
        {\textbf{0.13\,s} warm, $\leq$ SAT};
    \node[badge=cbad, minimum width=34mm, anchor=east] (bb1) at (au.east |- 0,0.65)
        {\textbf{11.6--22.2\,s} cold start};
    \node[badge=cbad, minimum width=34mm, anchor=east] (bb2) at (au.east |- 0,-0.25)
        {\textbf{OOM} at load on 8\,GB};
    \node[badge=cbad, minimum width=34mm, anchor=east] (bb3) at (au.east |- 0,-1.15)
        {\textbf{15--48\,s} compile / length};
    \node[mbox=cbad, inner sep=0pt,
          fit={(pr.west |- bb1.north) ($(bb1.west |- bb3.south)+(-3mm,0)$)}] (sa3) {};
    \node[align=center] at ($(sa3.center)+(0,4.6mm)$) {Stable Audio 3\\official PyTorch stack};
    \node at ($(sa3.center)+(0,-4.2mm)$) {\includegraphics[height=7.5mm]{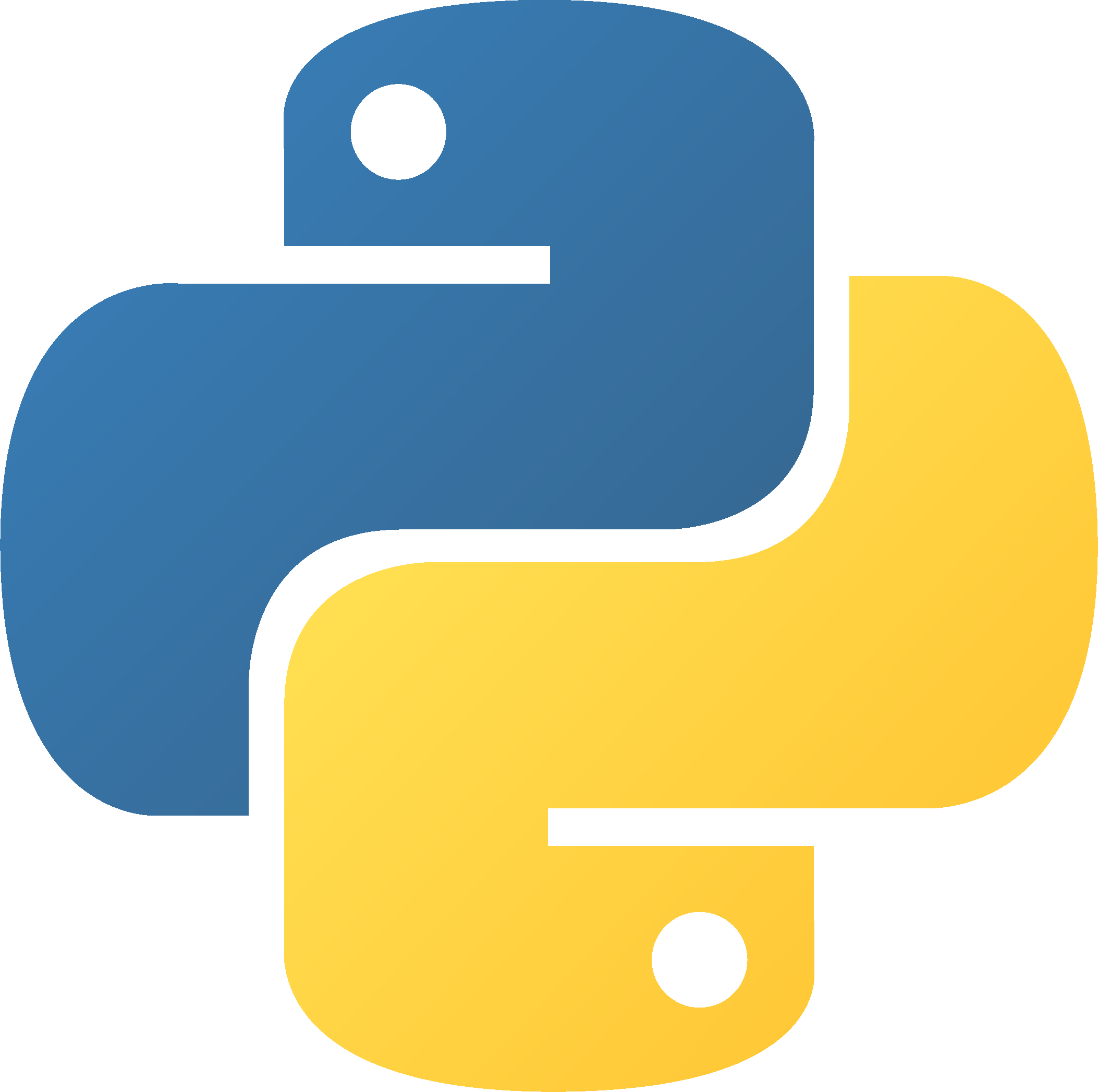}};
    \begin{scope}[on background layer]
      \node[panel=cgood, fit={($(pr.west)+(-2mm,0)$) ($(pr.north)+(0,4.5mm)$)
                              ($(gb1.south)+(0,-1.8mm)$) ($(au.east)+(2mm,0)$)}] (pbot) {};
      \node[panel=cbad, fit={($(sa3.north)+(0,4mm)$) ($(sa3.south)+(0,-1.8mm)$)
                             (sa3.center -| pbot.west) (sa3.center -| pbot.east)}] (ptop) {};
    \end{scope}
    \node[anchor=north west, font=\tiny\bfseries, text=cbad!85] at (ptop.north west)
        {\;rented / framework-bound};
    \node[anchor=north west, font=\tiny\bfseries, text=cgood!90] at (pbot.north west)
        {\;owned / one binary};
    \node[chip] (c2) at (ptop.south |- 0,-2.45) {CUDA graph + NEON};
    \node[chip, left=2.5mm of c2]  (c1) {C/CUDA, zero deps};
    \node[chip, right=2.5mm of c2] (c3) {windowed decode};
    \draw[->, black!55, line width=0.8pt] ($(ptop.south)+(0,-0.2mm)$)
        -- ($(c2.north)+(0,0.4mm)$);
    \draw[->, black!55, line width=0.8pt] ($(c2.south)+(0,-0.4mm)$)
        -- ($(pbot.north -| ptop.south)+(0,0.2mm)$);
    \node[font=\bfseries, text=cacc!90, anchor=east]
        at ($(ptop.south)!0.5!(c2.north)+(-1.4mm,-0.1mm)$) {\textit{aria}};
    \node[font=\tiny, text=black!60, anchor=west]
        at ($(ptop.south)!0.5!(c2.north)+(1.4mm,-0.1mm)$) {same checkpoints};
  \end{tikzpicture}
  \caption{With \textit{aria}, the same Stable Audio~3 checkpoints move from a
  framework-bound serving stack (\emph{top}) to a single dependency-free binary on hardware
  people own (\emph{bottom}): $7\times$ faster cold start, warm parity, and both model
  variants on a Raspberry~Pi~5. Owning the pipeline also makes activation steering a runtime
  feature, injected at \textcolor{cbad}{\textbf{1}} the DiT residual stream,
  \textcolor{cgood}{\textbf{2}} the latent, or \textcolor{cacc}{\textbf{3}} the text
  conditioning.}
  \label{fig:overview}
\end{figure}
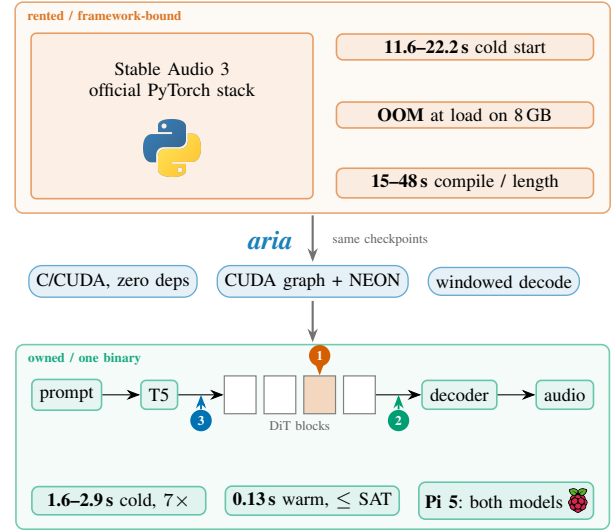

Local-inference work suggests a way forward, from the dependency-free
llama.cpp family~\cite{llamacpp2023} to portable engines such as ONNX
Runtime~\cite{onnxruntime} and DwarfStar's port of datacenter-scale language
models to user hardware~\cite{dwarfstar2025}. The engineering lesson is general:
once the serving stack is removed, model classes once treated as cloud-only can
become practical on owned hardware.

Audio generation is a particularly strong candidate for this shift. Open music
models are much smaller than frontier language models, roughly $1$--$10$\,B
parameters, and diffusion inference scales with denoising steps rather than the
far longer acoustic token sequences of autoregressive systems. The two leading
open-source music generators, ACE-Step~1.5~\cite{acestep2025} and Stable
Audio~\cite{sa3}, are both built on diffusion transformers (DiTs). We take Stable Audio~3 (SA3) as the
reference target and ask a systems question: how much of its deployment cost
belongs to the model itself, and how much belongs to the surrounding framework?

We answer that question with \textit{aria}, a dependency-free C/CUDA runtime for
SA3 that runs the full text-to-music pipeline, both model variants, on commodity
GPUs and CPU-only hardware (Figure~\ref{fig:overview}). The runtime executes the
tokenizer, text encoder, transformer denoiser, and autoencoder in a single native stack,
stores weights from half precision down to $8$- or $4$-bit integers, runs the arithmetic
in $8$-bit where it pays off, and holds long-form generation within bounded memory. On a Raspberry~Pi~5
(8\,GB), the small model generates a $10$\,s stereo clip at a $1.9$\,GB full-precision peak,
which quantization cuts to $0.84$\,GB at $8$-bit. The $1.2$\,B-parameter medium model
runs CPU-only at $4$-bit (${\sim}200$\,s; $0.9$\,GB resident, $3.6$\,GB peak).

A native runtime also owns every intermediate tensor, which makes activation
steering a runtime primitive rather than a Python-side patch or a retrained
adapter: directions can be injected into the generation graph at negligible cost,
giving semantic control while preserving the runtime's deployment advantages.

We use sonic seasoning, taste-conditioned music generation, as a case study for
this capability~\cite{spanio2025frontiers, spanio2024aixia}. Taste is a useful stress test because it is difficult to specify
lexically, grounded in an external crossmodal literature, and therefore
informative about whether a learned steering target captures genuine semantic
change or merely responds to degradation.

We make four contributions. (i)~\textbf{The \textit{aria} runtime}: a
dependency-free C/CUDA engine that runs SA3's full pipeline (text encoder,
transformer denoiser, autoencoder) on GPUs, CPU-only systems, and embedded-class
hardware, with a calibrated efficiency study against the official implementation
across warm latency, cold start, peak memory, and CPU-only generation.
(ii)~A \textbf{deployment quantization study}: weight storage from half precision down
to $4$-bit and an $8$-bit-arithmetic mode on GPU tensor cores and ARM, made
memory-frugal by releasing the full-precision weights once compressed. Rather than assume a
fidelity budget, we measure the quality cost with three independent checks (prompt adherence,
distributional quality, and taste preservation). On all three, 8-bit stays within re-seed noise, and
4-bit is sufficient, at a measurable but bounded cost, to put a $1.2$\,B model on an
$8$\,GB Pi.
(iii)~An \textbf{in-graph activation-steering interface} with zero measurable
overhead, bit-exact disable behaviour, and dose-response parity with the reference.
(iv)~A \textbf{bounded semantic-control case study} that uses sonic seasoning to
test a hard-to-lexicalize attribute under a multi-oracle protocol.
The result is a quantized native runtime with in-graph control, a
practical basis for on-device semantic audio generation.

\section{Related Work}
\label{sec:related}
\subsection{Generative audio models}
Text-to-music systems fall into two broad families: autoregressive transformers over discrete
acoustic tokens, and latent-diffusion models that denoise in the latent space of a neural audio
autoencoder. Diffusion cost scales with denoising steps rather than sequence length, and the leading
open-weight systems, ACE-Step~\cite{acestep2025} and the Stable Audio family, are DiT-based. Our
reference is Stable Audio~3 (SA3)~\cite{sa3}. It couples a SAME semantic-acoustic autoencoder~\cite{same2025},
a 256-dimensional continuous latent, to a diffusion-transformer (DiT) denoiser~\cite{dit}
conditioned on text via T5Gemma~\cite{gemma2024} cross-attention and trained with flow
matching~\cite{flowmatching}. We study its \emph{small-music} (20 DiT blocks,
$d_\mathrm{model}{=}1024$) and \emph{medium} (24 blocks, $d_\mathrm{model}{=}1536$) configurations. The DiT residual stream of per-block hidden states is the substrate that activation
steering reads and writes. The broader shift toward live generation, exemplified by
Magenta\,/\,Lyria~RealTime~\cite{magentart2025}, makes warm, resident runtimes more attractive than
cold per-request processes.

\subsection{Generative models on edge devices}
State-of-the-art generative audio increasingly runs behind closed APIs such as
Suno\footnote{\url{https://suno.com}, accessed 2026-07-04.} and
Udio\footnote{\url{https://www.udio.com}, accessed 2026-07-04.}. Open-weight
reference stacks, by contrast, are typically heavyweight Python pipelines that assume a discrete
GPU, and community CPU ports of SA3 remain tied to PyTorch~\cite{obsidian}. The ggml
lineage (llama.cpp~\cite{llamacpp2023}, whisper.cpp~\cite{whispercpp2022},
DwarfStar~\cite{dwarfstar2025}, stable-diffusion.cpp~\cite{sdcpp}) shows that a single,
self-contained C/C++ program with memory-mapped weights, low-bit quantization, and hand-written
vector and GPU kernels can deliver interactive inference on consumer hardware with near-zero cold start.
Portability-first engines such as ONNX Runtime~\cite{onnxruntime}, ExecuTorch~\cite{executorch},
Apple MLX~\cite{mlx2023}, and NVIDIA TensorRT~\cite{tensorrt} trade model-specific kernels for
generic graph execution. Neither approach has yet reached latent-diffusion music: to our knowledge,
no dependency-free C runtime exists for a state-of-the-art music model, a gap that \textit{aria}
fills.

\subsection{Quantization for on-device diffusion}
Low-bit inference is the lever that makes these runtimes fit budget hardware. Post-training
quantization (PTQ) compresses LLMs to 8-bit weights \emph{and} activations without
retraining~\cite{smoothquant}, diffusion-specific PTQ pushes image models to 4-bit
weights~\cite{qdiffusion}, and PTQ has recently been extended to audio diffusion
transformers~\cite{ptqaudiodit}. These works tune the quantizer for a fixed fidelity target. We instead address deployment. Our native
runtime treats precision as a first-class axis, spanning half-precision, $8$-bit, and $4$-bit storage
plus $8$-bit arithmetic. By releasing the full-precision weights once compressed, it makes lower
precision a memory \emph{replacement} rather than an addition. It ships hardware-specific $8$-bit
kernels (integer tensor cores on the GPU, the matching instruction on ARM), and gates each precision
against fp16 on independent objective checks of prompt adherence, distributional quality, and taste
preservation.

\subsection{Steering techniques}
A frozen generative model can be controlled either by editing its weights or by intervening on its
activations at inference time. The activation-based route rests on the linear representation
hypothesis, according to which interpretable attributes are encoded approximately as linear
directions in hidden space. Adding one vector to the residual stream can therefore shift an
attribute while leaving the rest of the computation intact. Activation addition~\cite{actadd} and
representation engineering~\cite{repe} extract such directions from contrastive examples and inject
them at generation time. Tan et al.~\cite{tan2024} characterize when these vectors generalize. This
paradigm is now well represented in audio and music~\cite{camporese_audioscope,facchiano2025,tada2026,musicrfm,singh2025,paek2025}.
We take our injection design from Camporese et al.~\cite{camporese_audioscope}. The weight-based
family (LoRA~\cite{lora}, ControlNet~\cite{controlnet}, concept sliders~\cite{conceptsliders}) can
provide strong control, but it requires training, per-attribute data, and a growing parameter
footprint. LoRA, the de-facto parameter-efficient standard, is our training-based comparator. Our
case study targets a gustatory attribute, the five basic tastes: ``sonic seasoning'' reports
reliable crossmodal associations between tastes and acoustic parameters~\cite{crisinel2010,mesz2011}.
They are stable enough to shift the perceived taste of food~\cite{knoferle2012,wang2015},
and Spence later synthesized them under the label of gastrophysics~\cite{spence_gastrophysics}. These
correspondences ground the reference used to assess steered generation; Section~\ref{sec:method}
gives the mapping.

\section{Methodology}
\label{sec:method}

\subsection{The aria runtime}
\label{sec:runtime}

\textit{aria} carries the dependency-free native-runtime discipline of Section \ref{sec:related} over
to audio latent diffusion. It targets the two settings in which the official Stable Audio~3 PyTorch
stack is least practical: interactive or live generation from a resident service~\cite{magentart2025},
and budget or CPU-only inference, where framework startup, GPU-context setup, and a multi-second cold
start dominate short generations. \textit{aria} is a single-purpose engine for SA3 in about
$7.7$k lines of C and CUDA, with no linear-algebra library, deep-learning framework, or other
third-party dependency beyond the C math library and a threading runtime.
It implements the whole pipeline behind a small dispatch layer: a text tokenizer, the T5Gemma text
encoder~\cite{gemma2024}, the diffusion transformer (DiT) that denoises the audio latent over eight
sampling steps, and the audio autoencoder~\cite{same2025} (a $256$-dimensional latent at
${\sim}10.7$\,Hz). Supporting a new architecture adds one file and a registry entry, not a kernel
rewrite. Weights are mapped directly from disk in half precision (fp16), with
full precision (fp32) kept only where fidelity is sensitive. Optional lower-precision storage, down
to $8$- or $4$-bit integers, keeps the medium model and small-memory GPUs within budget; once the
compressed copy exists the runtime releases the original, so a lower precision \emph{replaces} rather
than adds to the footprint, cutting the Pi's peak memory from $1.9$ to $0.84$\,GB at $8$-bit.

One set of operations runs behind two backends. The CPU path is vectorized and multi-threaded,
carries an $8$-bit-integer matrix-multiply variant, and doubles as the correctness reference and the
CPU-only production target. The GPU path uses half-precision tensor-core matrix multiplies and
attention, including a banded decoder kernel; its per-step denoising loop is captured once and
replayed as a single GPU graph. A windowed decoder caps memory at byte-identical output, and a
streaming variant decodes only the frames it emits ($3.1\times$ faster per chunk on the Pi, still
byte-identical); together they let the runtime generate and stream long-form audio on a
Raspberry~Pi~5. A resident batch mode and an HTTP server keep the model loaded across jobs for
multi-client serving, roughly doubling throughput ($0.60$ vs.\ $1.23$\,s per job) at identical output.

\paragraph{Steering interface.} Because the runtime owns every intermediate tensor, activation
steering becomes a built-in feature rather than a Python-side patch. \textit{aria} loads precomputed
direction vectors and applies them at three points in the pipeline, either adding the direction or
projecting onto it and optionally only during a chosen span of sampling steps: the transformer's
residual stream, the compact audio latent, and the text conditioning. This reproduces the reference
intervention exactly ($\delta=\alpha\lVert h\rVert\,\hat d$; Section \ref{sec:method}). On the GPU the
steering step lives inside the captured graph and reads a strength value refreshed each step (zero
outside the chosen span), so enabling it triggers no re-capture and adds no measurable overhead; at
strength $0$ the output is bit-identical to the base model. The taste case study of
Section \ref{sec:method} is thus a property of the runtime, not a fork of it.

\paragraph{Quantization and its evaluation.} The transformer's weights can be stored in half
precision, or compressed to $8$-bit or $4$-bit integers (labelled q8 and q4). An optional mode also
runs the arithmetic in $8$-bit, quantizing activations alongside weights (W8A8), on the GPU's integer
tensor cores or the equivalent ARM instruction. Because the compressed copy owns its memory, the
runtime frees the full-precision source once packed, so a lower precision \emph{reduces} resident
memory rather than adding to it. We gate
each precision empirically against the fp16 output on three independent objective checks
(Section \ref{sec:res-quant}, Table~\ref{tab:quant}) rather than assuming a fidelity budget.

\paragraph{Efficiency protocol.} To compare deployments fairly, we separate three regimes.
\emph{Warm} generation, with the model already resident in-process on both sides, is the cost a live
or batched service pays per request. \emph{Invocation} is a one-shot command-line call that
additionally re-pays per-process setup (text encoding and moving weights to the GPU). \emph{Cold
start} is process launch plus weight load plus one generation, the cost an edge endpoint pays on
first use. We report all three on an RTX~3070 GPU and a CPU-only tier (Table~\ref{tab:efficiency}),
analyzed in Section \ref{sec:res-runtime}.

\subsection{Taste steering and sonic seasoning}

\emph{Sonic seasoning} is the crossmodal finding that music systematically shifts perceived taste
(consonant high-pitched legato reads as \textsf{sweet}, dissonant low music as \textsf{bitter})
across the five basic-taste axes
$\{\textsf{sweet},\textsf{sour},\textsf{bitter},\textsf{salty},\textsf{spicy}\}$~\cite{crisinel2010,mesz2011,knoferle2012,wang2015}.
We ask whether a frozen music generator can be steered to \emph{produce} audio carrying these
associations. The machinery steers any difference-in-means direction. We choose taste as a difficult
attribute that remains externally measurable: continuous, weakly lexicalized, and grounded in an
independent psychophysics literature. We treat SA3 (Section \ref{sec:runtime}) as a black box and
intervene on its activations. The taste oracle we optimize is susceptible to metric gaming, so a
multi-oracle protocol distinguishes genuine control from degradation that artificially raises the
target.

\subsubsection{Direction and injection site}
Each DiT block writes its update additively, so its output lies in the same $d_{\text{model}}$ space
as its input. For axis $i$ and block $\ell$ we estimate a unit direction $d_i^{(\ell)}$ pointing from
``low-$i$'' to ``high-$i$'' by difference-in-means, the simplest estimator known to
transfer~\cite{actadd,repe,tan2024,camporese_audioscope}, over each contrastive set $S^{\pm}_i$:
\begin{equation}
\label{eq:diffmean}
d_i^{(\ell)} = \frac{\bar{h}^{+}_{i,\ell} - \bar{h}^{-}_{i,\ell}}{\bigl\lVert \bar{h}^{+}_{i,\ell} - \bar{h}^{-}_{i,\ell}\bigr\rVert_2},\qquad
\bar{h}^{\pm}_{i,\ell} = \frac{1}{|S^{\pm}_i|}\sum_{x\in S^{\pm}_i}\frac{1}{T}\sum_{t=1}^{T} h_\ell(x)_t.
\end{equation}
The contrast comes from two sources. \emph{Prompt-side} sets are caption pairs differing only in a
taste descriptor, self-contained but encoding only what the text encoder associates with that word.
\emph{Audio-side} sets ground the contrast in the reference dataset~\cite{normsonic}, $377$ music clips with per-clip basic-taste ratings. We rank these by rating on $i$ and feed the
top-/bottom-$k$ clips to SA3 as an initial audio input, capturing the residuals it produces on audio
humans actually heard as sweet or sour. Audio-side directions are more monotonic and more robust to
over-steering, so we use them by default while retaining prompt-side directions as a controlled
comparison.

\subsubsection{Additive injection}
At inference we add a scaled copy of the direction to the chosen block's output, for a single
strength $\alpha\ge0$,
\begin{equation}
\label{eq:inject}
h \;\leftarrow\; h + \alpha\,\lVert \bar{h}\rVert\,d_i^{(\ell)},
\end{equation}
where $\lVert\bar{h}\rVert$ is the mean baseline residual norm at $\ell$, so $\alpha$ is in units of
the typical residual magnitude, comparable across blocks and both model sizes. The intervention
touches one block, adds one rank-one term, and has one hyperparameter. Setting $\alpha=0$ leaves the
pass unchanged and gives the matched, seed-identical baseline. We sweep $\alpha$ up to $1.0$ because
the response is non-monotonic: a useful regime exists only at low strength, while larger $\alpha$
severely degrades the audio.

\subsubsection{Training-based baseline: per-axis LoRA}
To quantify what training-free steering sacrifices, we train a per-axis LoRA~\cite{lora} ($r=8$) on
the \emph{same} contrasts with the backbone frozen, using a trigger-token objective (high-$i$ clips
captioned ``\textsf{$i$} taste'' vs.\ a neutral ``music''). Unlike steering, it needs a per-axis
optimization run, adds served parameters, and cannot be removed by zeroing a scalar. Both arms share
the extraction data, pipeline, and evaluation, and each method's cost is logged
(Section \ref{sec:res-lora}).

\subsubsection{Multi-oracle evaluation}
The central risk is circularity: the quantity we optimize is also the quantity one might be tempted
to report as evidence of success. Our target is \emph{wav2taste}~\cite{wav2taste}, a learned
audio-to-taste regressor. The effect on axis $i$ is
$\Delta_i(\alpha)=\widehat\tau_i(x_\alpha)-\widehat\tau_i(x_0)$ against the matched $\alpha=0$
baseline. But wav2taste is imperfect (held-out macro Pearson $r\approx0.67$; best sweet $0.82$, worst sour $0.59$) and can over-score out-of-distribution audio. We therefore guard it with
three independent oracles: \textbf{CLAP}~\cite{clap,msclap} text--audio similarity to
``$X$-tasting music'' anchors, which shares no training signal, as a semantic cross-check;
\textbf{Fr\'echet Audio Distance (FAD)}~\cite{fad} over CLAP embeddings as a degradation detector;
and \textbf{audio drift}, $1-\cos(\phi(x_\alpha),\phi(x_0))$ on CLAP embeddings, as perturbation
magnitude. All clips are loudness-normalized to $-14$\,LUFS first, so no metric moves by loudness.
We treat a setting as \emph{genuine} only when $\Delta_i$ and CLAP rise together at low drift and
bounded FAD. When $\Delta_i$ rises while CLAP falls and FAD or drift increase, we report
degradation instead. A two-stage funnel locates this regime: a target-only per-layer scan
\emph{nominates} a candidate block per axis, then a dense $\alpha$-window at that block is
\emph{selected} by all four oracles. The nomination is target-only by design. It is a
cheap localizer, while the multi-oracle window is the actual gate, so the reported operating point
is never chosen by the target alone. That ordering also exposes the failure mode. The scan ranks the
final block first for the intense tastes, but the oracle panel then \emph{overturns} that block as
degradation (Section \ref{sec:res-layers}). A multi-oracle nomination would have hidden this rather than
surfaced it. The candidate block is retained only if it survives the panel. Otherwise its clean
operating point, or the axis, is rejected.

\section{Experiments and Results}
\label{sec:experiments}
\begin{figure}
  \centering
  \includegraphics[width=\linewidth]{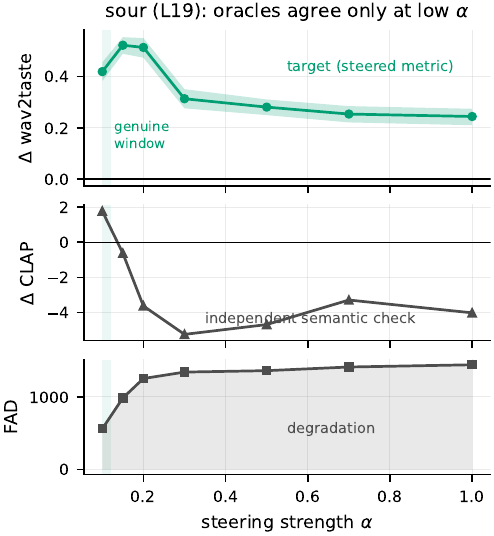}
  \caption{The degradation demonstration (\textsc{sour}, small-music, its best layer L19), as two
  stacked panels sharing $\alpha$ (no dual axis). \emph{Top:} the steering target (wav2taste $\Delta$,
  with a $95\%$ bootstrap CI band over the paired per-clip deltas) and the independent CLAP $\Delta$
  on one axis; both rise inside the shaded low-$\alpha$ genuine window, then diverge as wav2taste stays
  high while CLAP collapses. \emph{Bottom:} FAD explodes past the window. The high-$\alpha$ wav2taste
  gain is degradation that games the oracle. Colours are the fixed colourblind-safe taste palette
  (CLAP/FAD are shown neutral, being independent checks rather than tastes).}
  \label{fig:degradation}
\end{figure}

\begin{figure}
  \centering
  \includegraphics[width=\linewidth]{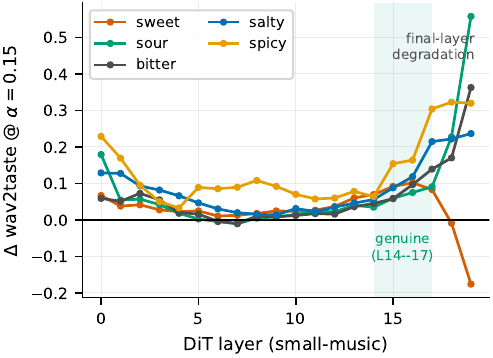}
  \caption{Every DiT layer steered individually at $\alpha{=}0.15$ (small-music; $n{=}15$ clips
  per point). \textsc{sweet} peaks mid-late (L16) and collapses at the final blocks, while the four
  intense tastes spike at the final layer, a ranking the quality control later overturns as
  degradation (their genuine site is the shaded L14--17). Medium behaves alike, peaking at
  L20--21.}
  \label{fig:layers}
\end{figure}


\begin{figure}
  \centering
  \includegraphics[width=0.95\linewidth]{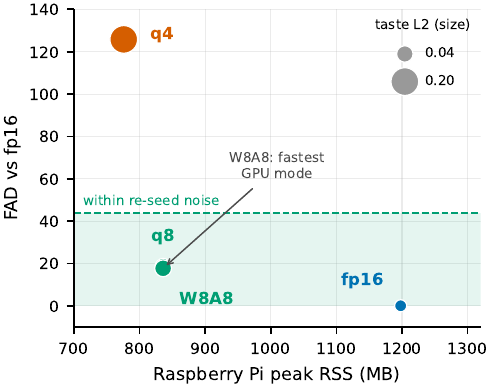}
  \caption{Quantization performance in one space (small-music; all GPU device-resident, fp16 decoder,
  so only DiT precision differs). Quantizing slides a model left (less Raspberry-Pi memory) at a
  fidelity cost (up: FAD vs.\ the fp16 set); marker area grows with the wav2taste taste-vector L2. The
  shaded band is the FAD a mere fp16 re-seed induces. 8-bit (q8, W8A8) stays inside the band and small,
  and within the $\Delta$CLAP and taste floors too (Table~\ref{tab:quant}), so it is indistinguishable
  from fp16; W8A8 is also the fastest GPU mode. 4-bit leaves the band for little extra memory but is
  what runs the $1.2$\,B medium model on an $8$\,GB Pi.}
  \label{fig:quant}
\end{figure}

\begin{table*}[t]
  \centering
  \small
  \caption{Runtime efficiency of \textit{aria} versus the official Stable Audio~3
  PyTorch implementation: $10$\,s clip, $8$ sampling steps, steering active, on an
  RTX~3070 (8\,GB) with an i9-10900KF (CPU-only rows use $20$ threads).
  \emph{Warm} = generation with the model already resident in-process (both sides);
  \emph{invocation} = a one-shot command-line call that re-pays per-process setup;
  \emph{cold} = process spawn $+$ weight load $+$ one generation. VRAM is the peak
  GPU process footprint; ``--'' where not applicable.}
  \label{tab:efficiency}
  \begin{tabular}{lcccc}
    \toprule
    & \multicolumn{2}{c}{small-music} & \multicolumn{2}{c}{medium} \\
    \cmidrule(lr){2-3}\cmidrule(lr){4-5}
    Configuration & Time (s) & VRAM (MB) & Time (s) & VRAM (MB) \\
    \midrule
    aria GPU (warm)       & \textbf{0.13} & \textbf{1395} & \textbf{0.37} & \textbf{4215} \\
    aria GPU (invocation) & 1.23          & --            & 2.55          & --            \\
    aria (cold start)     & \textbf{1.6}  & --            & \textbf{2.9}  & --            \\
    aria CPU-only         & 2.5           & --            & 9.8           & --            \\
    \midrule
    SA3 GPU (warm)        & 0.146         & 2330         & 0.443         & 5948         \\
    SA3 (cold start)      & 11.58         & --            & 22.23         & --            \\
    \bottomrule
  \end{tabular}
\end{table*}

\begin{table}
  \centering
  \caption{Precision ladder (small-music). Top: deployment cost (GPU warm $10$\,s and process VRAM on
  an RTX~3070, Raspberry~Pi~5 peak memory; MB). Bottom: fidelity vs.\ the fp16 reference on the three
  objective checks of Section \ref{sec:exp-quant} ($\Delta$CLAP prompt adherence; FAD distributional
  quality in a $32$-dimensional reduction of CLAP embeddings; $\Delta$taste, wav2taste $5$-D vector).
  Re-seed reference-noise floors: $\Delta$CLAP $\pm0.004$, FAD $44.0$, $\Delta$taste $0.153$. 8-bit
  (q8, W8A8) stays within every floor while cutting VRAM ${\sim}21\%$ and Pi memory to $0.84$\,GB, and
  W8A8 (8-bit arithmetic) is the fastest GPU mode; 4-bit crosses every floor but is what puts the
  $1.2$\,B medium model on an $8$\,GB Pi. fp16 reproduces fp32 ($r{=}1.00$).}
  \label{tab:quant}
  \small
  \setlength{\tabcolsep}{4pt}
  \begin{tabular}{lccccc}
    \toprule
              & fp32   & fp16   & q8      & W8A8            & q4       \\
    \midrule
    GPU\,(s)  & --     & $0.13$ & $0.20$  & \textbf{0.10}   & $0.22$   \\
    VRAM      & --     & $1510$ & $1190$  & $1190$          & $1128$   \\
    Pi\,mem   & $1912$ & $1198$ & $836$   & $836$           & $776$    \\
    \midrule
    $\Delta$CLAP   & --  & ref    & $+0.002$ & $-0.001$      & $-0.020$ \\
    FAD            & --  & ref    & $18.6$   & $17.7$        & $125.8$  \\
    $\Delta$taste  & --  & ref    & $0.035$  & $0.044$       & $0.199$  \\
    \bottomrule
  \end{tabular}
\end{table}

\begin{table}
  \caption{Dense-window multi-oracle steering (small-music, fp32), each axis at its
  wav2taste-best layer. \textsc{sweet} (L16) is shown across the full $\alpha$ sweep; the intense
  tastes (L19) at $\alpha{=}0.1$ vs.\ $0.15$. In each block the target ($\Delta$w2t) and the
  independent CLAP check rise together only at low $\alpha$ (the genuine window); past it CLAP flips
  negative (\textbf{bold}) and FAD jumps while wav2taste keeps climbing -- metric-gaming.
  \textsc{salty} barely moves CLAP even when clean; \textsc{spicy} is marginal.}
  \label{tab:window}
  \centering
  \small
  \begin{tabular}{lccccc}
    \toprule
    Axis & $\alpha$ & $\Delta$w2t & $\Delta$CLAP & FAD & drift \\
    \midrule
    \multicolumn{6}{l}{\emph{sweet (L16): full sweep}}\\
    sweet & $0.1$  & $+0.075$ & $+1.05$          & $73$   & $0.11$ \\
          & $0.15$ & $+0.097$ & $+1.43$          & $144$  & $0.16$ \\
          & $0.2$  & $+0.104$ & $+1.83$          & $274$  & $0.25$ \\
          & $0.3$  & $+0.119$ & $+2.11$          & $515$  & $0.39$ \\
          & $0.5$  & $+0.098$ & $+0.82$          & $777$  & $0.52$ \\
          & $0.7$  & $+0.056$ & $\mathbf{-1.16}$ & $933$  & $0.57$ \\
          & $1.0$  & $-0.025$ & $\mathbf{-3.88}$ & $1257$ & $0.72$ \\
    \midrule
    \multicolumn{6}{l}{\emph{intense (L19): genuine at $\alpha{=}0.1$}}\\
    sour   & $0.1$  & $+0.419$ & $+1.77$          & $561$  & $0.40$ \\
           & $0.15$ & $+0.522$ & $\mathbf{-0.63}$ & $985$  & $0.62$ \\
    bitter & $0.1$  & $+0.314$ & $+1.66$          & $836$  & $0.54$ \\
           & $0.15$ & $+0.360$ & $\mathbf{-1.15}$ & $1222$ & $0.73$ \\
    salty  & $0.1$  & $+0.209$ & $+0.32$          & $576$  & $0.41$ \\
           & $0.15$ & $+0.191$ & $\mathbf{-0.98}$ & $885$  & $0.57$ \\
    spicy  & $0.1$  & $+0.192$ & $-0.18$          & $301$  & $0.26$ \\
           & $0.15$ & $+0.267$ & $+0.25$          & $574$  & $0.42$ \\
    \bottomrule
  \end{tabular}
\end{table}

\begin{table}
  \caption{The \textit{aria} runtime reproduces the SA3/PyTorch steering dose--response: per-axis
  Pearson $r$ and MAE between aria and the reference on the wav2taste $\Delta$ (target) and CLAP
  $\Delta$ (independent check), over the shared per-$(\text{axis},\alpha)$ points ($n{=}35$ per
  model). aria tracks the target shape closely on both models; the looser small-model CLAP agreement
  is an fp16-vs-fp32 numerics offset that disappears on medium. \textsc{spicy} is
  noisiest on both.}
  \label{tab:parity}
  \centering
  \small
  \begin{tabular}{lcccc}
    \toprule
     & \multicolumn{2}{c}{wav2taste $\Delta$} & \multicolumn{2}{c}{CLAP $\Delta$} \\
    \cmidrule(lr){2-3}\cmidrule(lr){4-5}
    Axis & $r$ & MAE & $r$ & MAE \\
    \midrule
    \multicolumn{5}{l}{\emph{small-music}}\\
    sweet  & $0.984$ & $0.035$ & $0.979$ & $1.02$ \\
    sour   & $0.962$ & $0.040$ & $0.985$ & $1.42$ \\
    bitter & $0.979$ & $0.022$ & $0.973$ & $1.57$ \\
    salty  & $0.943$ & $0.047$ & $0.870$ & $2.00$ \\
    spicy  & $0.666$ & $0.067$ & $0.545$ & $4.29$ \\
    \textbf{pooled} & $\mathbf{0.953}$ & $\mathbf{0.042}$ & $\mathbf{0.859}$ & $\mathbf{2.06}$ \\
    \midrule
    \multicolumn{5}{l}{\emph{medium}}\\
    sweet  & $0.976$ & $0.033$ & $0.975$ & $0.46$ \\
    sour   & $0.949$ & $0.040$ & $0.921$ & $1.06$ \\
    bitter & $0.916$ & $0.037$ & $0.971$ & $1.40$ \\
    salty  & $0.833$ & $0.052$ & $0.934$ & $1.26$ \\
    spicy  & $0.614$ & $0.028$ & $0.956$ & $0.88$ \\
    \textbf{pooled} & $\mathbf{0.916}$ & $\mathbf{0.038}$ & $\mathbf{0.946}$ & $\mathbf{1.01}$ \\
    \bottomrule
  \end{tabular}
\end{table}

\begin{table*}
  \caption{Consolidated operating points for the injection-op, step-window, steering-site, and method
  ablations (small-music, fp32; $n{=}36$ clips per point, $12$ prompts $\times\,3$ seeds).
  $\Delta$w2t is the wav2taste target shift vs.\ the matched $\alpha{=}0$ baseline; $\Delta$CLAP is the
  independent semantic check (\textbf{bold} once it has flipped negative, i.e.\ off the genuine window);
  FAD is degradation vs.\ baseline. The four groups are read in the steering results
  (Section \ref{sec:results}).}
  \label{tab:ablate}
  \centering
  \footnotesize
  \begin{tabular}{llccccc}
    \toprule
    Axis & Method / site & Op-point & $\Delta$w2t & $\Delta$CLAP & FAD \\
    \midrule
    \multicolumn{6}{l}{\emph{Injection op (DiT residual)}}\\
    sour  & additive   & $\alpha{=}0.1$  & $+0.419$ & $+1.77$          & $561$  \\
          & additive   & $\alpha{=}0.15$ & $+0.522$ & $\mathbf{-0.63}$ & $985$  \\
          & projection & $\beta{=}1$     & $+0.492$ & $+0.60$          & $544$  \\
    sweet & additive   & $\alpha{=}0.3$  & $+0.119$ & $+2.11$          & $515$  \\
    \midrule
    \multicolumn{6}{l}{\emph{Denoise-step window (DiT residual, sour)}}\\
    sour  & early (steps 0--3) & $\alpha{=}0.2$ & $+0.530$ & $\mathbf{-4.26}$ & $1294$ \\
          & late (steps 4--7)  & $\alpha{=}0.2$ & $+0.515$ & $+0.80$          & $565$  \\
    \midrule
    \multicolumn{6}{l}{\emph{Steering site}}\\
    sour  & latent (256-D) & scale $1$   & $+0.385$ & $+1.34$          & $365$ \\
    sweet & latent (256-D) & scale $0.5$ & $+0.050$ & $+0.69$          & $197$ \\
    sour  & text (768-D)   & scale $0.5$ & $+0.299$ & $\mathbf{-3.04}$ & $935$ \\
    \midrule
    \multicolumn{6}{l}{\emph{Method: LoRA vs.\ best steering}}\\
    sour  & LoRA     & $\alpha{=}1.0$      & $+0.341$ & $\mathbf{-0.92}$ & $767$ \\
    sweet & LoRA     & $\alpha{=}0.5$      & $+0.032$ & $\mathbf{-0.10}$ & $171$ \\
    sour  & steering & proj.\ $\beta{=}1$  & $+0.492$ & $+0.60$          & $544$ \\
    sweet & steering & add.\ $\alpha{=}0.3$ & $+0.119$ & $+2.11$         & $515$ \\
    \bottomrule
  \end{tabular}
\end{table*}

\subsection{Experimental setup}

All conditions use $10$\,s clips, loudness-normalized to $-14$\,LUFS before scoring, with three
seeds $\{0,1,2\}$ and means reported. Steering strength $\alpha$ is expressed in units of the patched
block's mean residual norm (Section \ref{sec:method}). Directions are grounded in
\emph{norm-sonic-seasoning}~\cite{normsonic} ($377$ human-rated clips). We steer two released
configurations: small-music (fp32, $20$ blocks, $d_\text{model}{=}1024$) and medium
(fp16, $24$ blocks, $d_\text{model}{=}1536$). Medium in fp32 exceeds the $8$\,GB
reference GPU. Precision therefore differs with size. To keep the two apart, we re-run
small-music in \emph{both} fp32 and fp16 (Section \ref{sec:res-quant}). The two
dose--responses coincide ($r{=}1.00$), which isolates the medium--small differences as scale rather
than precision. Per-point significance is assessed with a
two-sided Wilcoxon signed-rank test on the $n{=}36$ paired deltas, Holm-corrected across the five
axes.

\paragraph{Extraction source and layer scan}\label{sec:exp-layers}
We compare prompt- vs.\ audio-side directions by dose-response monotonicity (Spearman $\rho$) and
over-steer robustness, holding block, grid, and seeds fixed. To localize \emph{where} taste is
steerable, a logistic probe identifies the layers at which the axes are linearly separable. We then
steer every DiT block at $\alpha=0.15$ ($5$ prompts $\times\,3$ seeds). Blocks are ranked by the
target metric alone. One candidate block per axis is then submitted to the oracle panel for
confirmation or rejection.

\paragraph{Dense window and oracle panel}\label{sec:exp-oracles}
At each axis's best layer on small-music we sweep $\alpha\in\{0.1,0.15,0.2,0.3,0.5,0.7,1.0\}$ over
$12$ prompts $\times\,3$ seeds, scoring every clip with the four-oracle panel of Section \ref{sec:method}
(wav2taste target, CLAP cross-check, FAD, drift). A model-size study repeats the \textsf{sweet}
sweep on medium, with each model evaluated at its own re-scanned best layer.

\paragraph{Ablation designs}\label{sec:exp-ablate}
Three ablation families reuse the dense-window panel on small-music. \emph{Injection op}: besides
additive steering we test projection, amplifying ($\beta{>}0$) or removing ($\beta{<}0$) the
direction component already present per token, swept both ways. \emph{Step window}: injection
confined to the early ($0$--$3$) vs.\ late ($4$--$7$) denoise steps at matched strength.
\emph{Steering site}: the same difference-in-means recipe applied to the compact audio latent and the
pooled text embedding, comparing residual, latent, and text injection. The training
comparator is the per-axis rank-$8$ LoRA of Section \ref{sec:method} ($800$ steps), dose-swept and scored
with the same panel.

\paragraph{aria reproduction and efficiency}
To confirm the runtime preserves steering behaviour, we re-run the dense window through
\textit{aria} with the identical exported direction vectors and $\alpha$ grid, and report
per-axis Pearson $r$ and MAE against the SA3/PyTorch reference on the wav2taste and CLAP
dose-responses. Efficiency is measured on the RTX~3070 and a CPU-only x86 tier under the warm /
invocation / cold protocol of Section \ref{sec:runtime} (Table~\ref{tab:efficiency}). The baseline is the
official Stable Audio~3 implementation (commit \texttt{dedace19}, 2026-07-01; torch~$2.7.1$, CUDA~12.6),
run in half precision with its native $8$-step sampler and no classifier-free guidance. With the
faster attention kernel unavailable, the repository's fallback attention path is active, and reported
memory is the process's GPU footprint. The repository disables several standard performance options by
default; we measured both that default and a tuned configuration (those options enabled, plus
compiling the transformer) and cite the \emph{faster} one throughout.

\paragraph{Quantization fidelity}\label{sec:exp-quant}
We characterize each precision against fp16 rather than fixing a fidelity budget. Every clip is
generated on the GPU with the decoder held at half precision, so a comparison isolates the
transformer's weight (q8, q4) or weight-plus-activation (W8A8) quantization from any device effect. We
use $24$ genre-diverse music prompts $\times\,3$ seeds ($72$ clips per precision, $10$\,s,
loudness-normalized). We score three metrics drawn from two independent audio models. Two come from
CLAP~\cite{msclap}, a text--audio embedding model: \emph{prompt adherence} (how closely each clip
matches its own text prompt) and \emph{distributional quality} (Fr\'echet Audio Distance, FAD~\cite{fad},
between the precision's clips and the fp16 set in a $32$-dimensional embedding space, reduced for
stability at this sample size). The third, from the separate wav2taste~\cite{wav2taste} model, is
\emph{taste preservation} (the distance between the $5$-D taste vectors of each clip and its fp16
counterpart). Re-seeding fp16 (seeds $3$--$5$) gives each metric a
reference-noise floor, the change a re-seed alone induces; a precision is within noise below it.



\subsection{Runtime benchmarks}
\label{sec:res-runtime}

Table~\ref{tab:efficiency} reports the three deployment regimes across both tiers. In the warm setting, with the model resident on both sides, \textit{aria} slightly edges the official implementation after a round of kernel tuning and request reuse. A steered $10$\,s clip runs in $0.13$ vs.\ $0.146$\,s (small-music) and $0.37$ vs.\ $0.443$\,s (medium). It cold-starts
$7.2$--$7.7\times$ faster and holds a $1.4$--$1.7\times$ smaller peak GPU-memory footprint
(Table~\ref{tab:efficiency}), and its resident batch mode amortizes per-process setup to $0.60$\,s
per job. The runtime also operates without a GPU. After a pass that parallelizes and vectorizes the
elementwise work between the matrix multiplies, a $10$\,s clip decodes in $2.5$\,s on
small-music (${\sim}4\times$ real time) and $9.8$\,s on medium on a
$20$-thread CPU, and chunked sliding-window generation streams at a real-time factor of
$0.6$--$4\times$ on the RTX~3070. A calibrated fast preset (six denoising steps instead of eight)
trims a further ${\sim}25\%$ off every transformer-bound path, with a taste-shift below the
seed-to-seed noise floor on both tiers.

At $60$\,s on GPU the picture inverts on the medium tier. A half-precision attention output stage and
running the mapping convolution as a matrix multiply put \textit{aria} at $1.28$\,s against the
reference's fused-attention path at $1.38$\,s, faster at equal precision. An opt-in $8$-bit-arithmetic
mode (W8A8, of $4$-bit-class fidelity) extends the lead to $1.12$\,s. Only on small-music, where
self-attention dominates the smaller autoencoder, does the official path stay ahead ($0.52$ vs.\
$0.38$\,s, $1.37\times$). CPU favours \textit{aria} too: its banded decoder renders $60$\,s of medium
$1.68\times$ faster ($48.0$ vs.\ $80.65$\,s), and the small-tier CPU gap to the PyTorch baseline
essentially closes ($1.09\times$). The official warm figure relies on the reference's kernel compiler,
which recompiles for \emph{each} new clip length (a $14.5$--$48.1$\,s penalty, poorly suited to
varying lengths); its medium loader also briefly needs ${\sim}7.1$\,GB, overflowing an $8$\,GB card,
so the benchmark used a low-memory load path.

\paragraph{Quantization is close to free at 8-bit, and enables the edge}\label{sec:res-quant}
Precision is a deployment axis, so we characterize rather than fix it (Table~\ref{tab:quant},
Figure~\ref{fig:quant}) with the three checks of Section \ref{sec:exp-quant}. To avoid validating
quantization with the same oracle the steering study optimizes, we lead with the two msclap-based
metrics that share no training path with the taste oracle: prompt adherence and FAD. We keep the
network-independent wav2taste L2 as a narrow corroborator, and scale each metric by the change a fp16 re-seed
induces. Eight-bit shows no measurable degradation on any of the three. Both q8 and W8A8 stay inside every
re-seed floor while cutting GPU memory ${\sim}21\%$ and Pi resident memory $2.3\times$, and W8A8 is the
fastest GPU mode ($0.10$\,s warm, on the GPU's integer tensor cores). Four-bit measurably crosses every
floor, most on prompt adherence ($5\times$ its floor). It nonetheless lets the $1.2$\,B medium model run
on the $8$\,GB Pi (${\sim}200$\,s); fidelity here is small-music, and the medium case is a memory-fit
result. fp16 reproduces the fp32 dose--response ($r{=}1.00$, MAE $0.003$; steering sweep), so precision
is not confounded with the model-scale results below.

\paragraph{Live steering}
Because steering runs inside the graph and the model stays warm, the control can be turned on during
a stream. With the \textsc{sweet} scale ramped $0{\to}0.5{\to}0$ across a 12-chunk stream, the
per-chunk taste score tracks the schedule (Spearman $\rho{=}0.78$, $p{=}0.003$), lagging on the
descent as the continuation inherits context. On the Raspberry~Pi~5, a steered clip costs the same as
an unsteered one ($32.9$ vs.\ $32.8$\,s), so semantic control remains fully on-device.

\subsection{Steering results}
\label{sec:results}

\emph{Extraction source.} On small-music, the prompt-derived \textsc{sweet} direction yields an
inverted-U response, peaking at $\Delta_{\text{w2t}}{=}{+}0.05$
($\alpha{=}0.1$) then collapsing to $-0.08$ at $\alpha{=}0.2$; the audio-derived direction reaches
the same peak but remains monotonic (Spearman $\rho{=}{+}0.94$) and still positive at $\alpha{=}0.2$.
We therefore use audio-side directions throughout.

\paragraph{A narrow clean window}\label{sec:res-degradation}
The central result is that wav2taste alone is not a reliable guide. Sweeping $\alpha$ at the
\textsc{sour} axis's best layer L19 (Figure~\ref{fig:degradation}, Table~\ref{tab:window}), the
oracles agree only at low strength. At $\alpha{=}0.1$ the shift is large
($\Delta_{\text{w2t}}{=}{+}0.42$, $p{=}2.9{\times}10^{-11}$, $d{=}3.71$) with CLAP positive, genuine
steering. Past $\alpha{=}0.1$ they diverge: wav2taste keeps climbing to its maximum at
$\alpha{=}0.15$ while CLAP flips negative and FAD jumps. By $\alpha{=}1.0$ the audio has
collapsed into noise-like output that wav2taste, weakest on sour ($r{\approx}0.59$), misreads as intense taste.
The metric's maximum lies \emph{inside} the degradation regime. In the clean window, \textsc{sweet},
\textsc{sour}, and \textsc{bitter} steer genuinely with CLAP confirmation, whereas \textsc{salty} and
\textsc{spicy} stay weak even when clean; we therefore claim usable control only for the first three
axes and report the case study as a bounded proof of concept, not broad taste controllability. On ten held-out prompts never used
for layer or strength selection, both operating points replicate (\textsc{sweet} $+0.164$,
\textsc{sour} $+0.508$; Holm $p{=}3.7{\times}10^{-9}$). An \emph{audio}-anchored CLAP variant
(similarity to the centroid of the top-taste clips instead of a text anchor) reproduces the same
boundary (\textsc{sour} $+0.12$ at $\alpha{=}0.1$, negative from $\alpha{=}0.3$). The window is
therefore neither a layer-selection artefact nor a text-anchoring artefact.

\paragraph{Layer selection depends on the attribute}\label{sec:res-layers}
Linear separability does not imply steerability. A logistic probe places taste in early text-encoding layers. Yet
steering each block at $\alpha{=}0.15$ (Figure~\ref{fig:layers}) shows \textsc{sweet} peaks mid-late
at L16 and collapses at the final blocks, while the intense tastes rank the \emph{final} block
first. Independent quality checks overturn that ranking: the scan-winning values already sit in the
degradation regime with CLAP negative, so layer selection cannot be delegated to the target. The
most reliable operating point is \textsc{sweet} at L16, a modest $\Delta_{\text{w2t}}{=}{+}0.119$ at
$\alpha{=}0.3$ ($p{=}3.2{\times}10^{-9}$, $d{=}1.20$) with CLAP $+2.11$ at bounded FAD
(Table~\ref{tab:window}).

\paragraph{Model scale}
A larger backbone widens the clean operating window rather than merely inflating the target metric.
For \textsc{sweet},
medium reaches a higher clean peak ($\Delta_{\text{w2t}}{=}{+}0.143$ at $\alpha{=}0.15$;
$p{=}8.9{\times}10^{-9}$, $d{=}1.32$) and holds a positive CLAP plateau farther out than
small-music. At $\alpha{=}0.1$ medium also degrades far less on the intense tastes (lower FAD
across axes) while keeping CLAP positive. Re-running the small model's full window in fp16
reproduces its fp32 dose--response (pooled $r{=}1.00$, MAE $0.003$ over all five axes), so
the medium--small differences are scale, not precision.

\paragraph{aria reproduces the dose--response}\label{sec:res-parity}
Re-running the dense window through \textit{aria} reproduces the reference closely
(Table~\ref{tab:parity}): pooled wav2taste $r{=}0.95$ (small) and $0.92$
(medium), with \textsc{sweet} tightest. CLAP agrees on medium ($r{=}0.95$) but is looser on small
($r{=}0.86$), reflecting a numerical offset (aria runs in half precision where the small reference
runs in full) that vanishes on medium. Both the genuine and degradation regimes carry over, so the
case-study conclusions hold.

\paragraph{Ablations: injection operator and step window}\label{sec:res-ablate}
Two ablations run entirely through the runtime's batch mode (the PyTorch injector supports
neither), consolidated in Table~\ref{tab:ablate}. \emph{Op}: projection \emph{amplification}
(scaling up the direction already present in each token) matches additive steering's peak \textsc{sour} shift at
markedly lower FAD and with CLAP still positive ($\beta{=}1$: $p{=}2.9{\times}10^{-11}$,
$d{=}4.08$), but it does nothing for \textsc{sweet}, whose residual component is too small to amplify.
The two operators are complementary across axes. Steering \emph{away} from the direction has little
effect on either axis.
\emph{Window}: confining injection to the \emph{late} denoise steps (4--7) roughly halves the
damage at matched taste shift (late $\alpha{=}0.2$: $p{=}2.9{\times}10^{-11}$, $d{=}5.10$, CLAP
positive, vs.\ the early steps' CLAP collapse), consistent with early steps setting global
structure while late steps shape the timbre both oracles read.

\paragraph{Steering sites vary by attribute}
\label{sec:res-sites}
The same protocol evaluates the pipeline's three injection sites (Table~\ref{tab:ablate}). The
compact-\emph{latent} site steers \textsc{sour} cleanly, a better quality-per-shift point than the
residual site (scale $1$: $p{=}2.9{\times}10^{-11}$, $d{=}3.11$, CLAP positive at low FAD), but fails
\textsc{sweet}, which inverts. The \emph{text-embedding} site fails both axes at
\emph{every} scale tested ($0.5$--$32$): even the smallest nudge reads as degradation and
\textsc{sweet} turns negative. This suggests that a mean shift in the text-embedding space moves the
conditioning off the data manifold rather than encoding ``more taste'', and corroborates the
prompt-side extraction failure at the embedding level. \textsc{sour}, an overtly timbral attribute, is steerable at latent and residual
sites alike, while \textsc{sweet} responds only to mid-late residual injection.

\paragraph{Steering versus LoRA}\label{sec:res-lora}
Under the same evaluation panel, the trigger-token LoRA (rank~$8$, $800$ steps) never reaches a
CLAP-positive operating point on either axis (Table~\ref{tab:ablate}): \textsc{sour} peaks with CLAP
negative, below steering's clean shift, and \textsc{sweet} barely moves ($p{=}0.16$, n.s.), turning
\emph{negative} at higher strength. A larger, longer-trained adapter (rank~$32$, $3\times$ the steps)
lands in the same place. The cost comparison is one-sided too: LoRA needs ${\sim}5$\,min of training,
$5.15$\,M parameters, and $10.4$\,MB per axis, whereas steering trains nothing, stores a $4$--$6$\,KB
vector, runs in-graph at no overhead, and is removable at serve time. For this lexically ill-defined
attribute, parameter updates add no control over a simple steering vector: steering wins on effect
and cleanliness, not only on cost.

\section{Discussion}
\label{sec:discussion}
\subsection{Implications of warm-parity edge deployment}

On warm throughput the two stacks are within a few percent, with \textit{aria} marginally ahead:
the runtime gains less from faster arithmetic than from removing framework startup, GPU-context
setup, and per-length compilation, the costs that dominate short generations. Interactive and
embedded settings benefit most from a resident, warm model, since cold start, memory footprint, and
predictable behaviour across clip lengths matter more there than steady-state speed. The one
remaining GPU deficit is long-form generation on the \emph{small} tier; on medium \textit{aria} overtakes the official path at exact precision, and CPU-only it renders the same clip faster.
Quantization is what makes the embedded tier practical: 8-bit shows no measurable degradation on any
of the three checks, releasing the full-precision copy turns lower precision into a memory
\emph{reduction} that reaches an $8$\,GB Pi at $4$-bit, and on the GPU 8-bit arithmetic is the
fastest mode, so compression and speed do not trade off.

\subsection{Learned oracles require independent checks}

When an intervention is tuned to maximize a learned metric, that metric ceases to be a faithful
indicator once the intervention becomes too aggressive: the target keeps rising while causally
independent checks collapse. The same limitation applies to layer selection, because the target can
mistake degradation for success. Both the strength and the site of an intervention must therefore
be supervised by quality signals that share no training path with the target. This caution extends
to any representation-editing evaluation that reports the optimized quantity as its own evidence.

\subsection{Scope and limitations}

The systems contributions (the runtime, the quantization study, the efficiency comparison) rest
on direct measurement and stand on their own. The steering study is scoped as a bounded case study. It intervenes on a single model family with
learned or distributional oracles, so we claim genuine control only for the three attributes where
independent checks confirm it and treat the rest as a negative result. Those operating points
replicate on held-out prompts, survive an audio-anchored re-test, and are stable across a precision
change (half versus full), the runtime port, and a stronger adapter (Section \ref{sec:results}).
What remains open is perceptual confirmation, since no automatic oracle equals human listening. A
listening study is therefore the natural next step rather than a gap in the present claims, which are
pitched at exactly what the oracles can and cannot certify.

\section{Conclusion}
\label{sec:conclusion}
We presented \textit{aria}, a dependency-free C/CUDA runtime for Stable Audio~3, to show that a
state-of-the-art latent-diffusion music model can be deployed without a datacenter-oriented serving
stack. In roughly $7.7$k lines of C, with no framework or third-party dependencies, aria runs
the full text-to-music pipeline on a commodity CPU and an inexpensive GPU: it matches the official
implementation's warm GPU throughput (slightly exceeding it after kernel tuning), cuts cold
start by about $7\times$, lowers peak GPU memory, and runs CPU-only at $\sim$$4\times$ real time, with
long-form GPU generation on the small tier its one remaining limitation. Treating precision as a
deployment axis, 8-bit weights leave prompt adherence, distributional quality, and taste all within
re-seed noise, 8-bit arithmetic runs fastest on the GPU, and releasing the full-precision weights once
compressed puts the $1.2$\,B model on an $8$\,GB Pi at $4$-bit.
Its bit-exact activation-steering interface reproduces the reference dose--response, and the
sonic-seasoning case study shows that taste control is real but confined to a narrow operating window
for a subset of attributes. The wider lesson is that semantic control should be evaluated with
independent quality checks, not with the optimized metric alone.

\subsection{Outlook}

Several extensions are clear. Fusing the decoder's kernels has reversed the
medium-tier long-form gap, so \textit{aria} overtakes the official path there at exact precision.
The remaining small-tier distance comes from the transformer's full self-attention, whose attention
weights we measure to be too spread out for a banded approximation; a fused attention kernel
(FlashAttention-class) behind a build flag is the open lever. A persistent server mode would keep the model resident
for multi-client serving and support latent-domain streaming continuation~\cite{magentart2025}. The
main open validation is perceptual. As future work we plan a human listening study, using pairwise
comparisons fit with a Bradley--Terry model~\cite{bradleyterry}, to externally confirm the
automatic oracles on the attributes where they indicate genuine control. Both SA3 variants run on a
Raspberry~Pi~5, the medium only at $4$-bit, where the $8$-bit-arithmetic path is the
fastest CPU mode.

\section*{Acknowledgment}
This work was funded by the European Union - NextGenerationEU, under the National Recovery and Resilience Plan (PNRR).

\bibliographystyle{IEEEtran}
\bibliography{references}

\end{document}